# An empirical dependence of frequency in the oscillatory sorption of $H_2$ and $D_2$ in Pd on the first ionization potential of noble gases[1]

Erwin Lalik

*Jerzy Haber Institute of Catalysis and Surface Chemistry, Polish Academy of Science, ul. Niezapominajek 8, Krakow 30-239, Poland; E-mail: nclalik@cyf-kr.edu.pl, tel: +48126395189, fax: +48124251923*

**Abstract.** Oscillatory heat evolution in sorption of $H_2$ and $D_2$ in Pd can be induced by an admixture of ca. 10 % vol. of an inert gas (He, Ne, Ar, Kr or $N_2$) to either isotope prior to its contact with palladium powder. The oscillations are represented in the form of a calorimetric time series, recorded using a gas flow-through microcalorimeter at the temperatures of 75 °C for $D_2$ and 106 °C for $H_2$. For both $D_2$ and $H_2$, the oscillation parameters change as a function of the kind of inert gas used: the amplitude increases and the frequency decreases in passing from He to Kr. An empirical dependence of the oscillation frequencies observed for various admixtures and normalized with respect to Kr has been found. Accordingly, the frequency is a function of a product of the first ionization potential and the square root of atomic mass of the inert gas (He, Ne, Ar, Kr or $N_2$). On the other hand, invariance of the thermal effects of sorption is evident from the integrated areas under the calorimetric time series yielding the molar heats of sorption conserved, irrespective of the inert gas admixture. A novel calibration procedure has been devised in order to deal with an instability of calibration factor arising in desorption of $H_2$ and $D_2$ from Pd. A method of dynamic calibration factor made it possible to obtain a good agreement between the heats of sorption and desorption of both $H_2$ and $D_2$ within individual sorption-desorption cycles for all inert gas admixtures.

## 1. Introduction

In a previous article[1] we reported a finding that the presence of ca. 10% of inert gas ($N_2$) in a gaseous mixture with hydrogen (protium) resulted in oscillations during the reaction of such mixture with metallic palladium at 106 °C. The process was carried out in a flow mode, monitored *in situ* using a gas flow-through microcalorimeter. Apart from the inert $N_2$ admixture, an initial lowering of the system pressure from the ambient to ca 600 hPa has also been found to be a necessary condition for the oscillations to develop. The oscillatory character is shown by the rate of heat evolution accompanying the reaction, the so-called thermokinetic oscillations. It has been confirmed experimentally, that the oscillations are a kinetic phenomenon, the total amount of heat generated during reaction remains constant, which means that the thermodynamic parameters of the process are not altered.[1]

In the previous experiments,[1] only a range of $H_2/N_2$ mixtures have been studied. Here we describe the results of further oscillatory experiments now extended not only by including the sorption of $D_2$ in palladium, but also by using the noble gases, He, Ne, Ar and Kr, apart from $N_2$, as inert admixtures with both isotopes. It has been found that the oscillatory kinetics in the sorption of deuterium in metallic palladium occurs under conditions similar to those applied for $H_2$ and, to our best knowledge, this is the first time that oscillatory sorption of deuterium in Pd is being reported. For both isotopes, using noble gases as inert admixtures resulted in oscillation intensity consistently increasing in the order He < Ne < Ar < Kr. Indeed, a functional relation has been revealed between the nature of the noble gas used as admixture and the parameters of the ensuing oscillations, based on a total of 113 experimental sorption-desorption cycles. This seems to suggest that the noble gases play a role in the mechanism that triggers the oscillations.

Noble gases are widely used in laboratories as the inert carriers diluting the actual sorbate in various experiments involving the solid–gas interactions. Usually it is assumed that their presence does not significantly alter the adsorption process they are used in, because of their chemical inactivity, although it is well recognized that they may undergo a weak, physical adsorption on various surfaces. Nevertheless, a handful of reports are now available showing that certain noble gases can play a more active role in adsorption as well as in heterogeneous catalysis. Thus it has been found that merely the presence of argon during pretreatment of certain types of gold supported catalysts can improve their performance in the process of epoxidation[2] and in the low temperature oxidation of $CO$.[3] The presence of helium or neon can strongly alter sorption capacity of the type A zeolites for water.[4,5,6] Likewise, both argon and helium were found to change the rate as well as the equilibrium of adsorption of various $C_4$ alkanes in the zeolites BEA, MFI and FER, with the effects consistently stronger for Ar than for He[7,8,9]. The difference in strength between the respective actions of Ar and He indicates that the kind of noble gas used in those experiments was consequential. This seems to be in agreement with a theoretical study using DFT, showing a clear trend in adsorptive interactions of He, Ne, Ar, Kr, and Xe with the Pd(111) surface.[10] The experiments described in the presented work provide a strong evidence of the possibility for sorptive reactions to be influenced by the presence of inert gases. They also confirm that an extent of such a "noble" intervention can be functionally related to the kind of noble gas being used. At the same time, the presented data also confirm that the heat of sorption and desorption is not affected by the presence of the inert gas, showing thereby that the effect of the noble gas in this case is clearly limited to altering the kinetics of sorption, without changing the reaction thermodynamics.

## 2. Experimental

**Materials.** Three kinds of powdered Pd samples of different granularity were used. The first will be referred to as the palladium powder (purity 99.999%, particle size $0.25 - 2.36$ mm), the second as the fine grained Pd powder of granularity less than 75 µm, both supplied by Aldrich Co., and the third as palladium sponge (44 µm – 149 µm) supplied by Johnson-Matthey. The following gases, provided by Linde Gas Poland S. A., were used: nitrogen (99.999%), hydrogen (99.999%), deuterium (99.999%), helium (99.999%), neon (99.995%), argon (99.999%) and krypton (99.99%).

**Equipment.** A Microscal flow-through microcalorimeter, Model FMC-4110 designed for use in isothermal mode at temperatures up to 240 °C and pressures up to 5 MPa has been applied. The design and operation of this microcalorimeter has been previously described in detail,[6] and its use has been reviewed elsewhere.[11,12] The instrument measures the rate of heat evolution accompanying a solid–gas interaction. A sample of sorbent is placed in a small microcalorimetric cell (7 mm in diameter) and the measurement is carried out in a flow-through mode. The cell is placed centrally within a larger, metal block acting as a heat sink, which ensures a steady removal of total evolving heat, thus preventing its accumulation within the cell. As a reaction is running within the cell, a minute difference of temperatures, between the vicinity of the cell and the locations closer to the outer edge of heat sink, can be measured continuously by a system of thermistors, appropriately located within the block. This signal can be used for a calculation of rates of heat evolution with application of a calibration factor (CF). For the latter to be extracted, each experiment (or a group of experiments) needs to be preceded (or followed immediately) by an *in situ* calibration pulse of controlled power and duration. The pulse is produced by a small electric coil sealed in the calibrator which is located axially within the calorimetric cell and totally surrounded by powdered sample. Either the peak height or area under the peak recorded on such a calibrating pulse is then used as a standard to calculate the calibration factor for individual experiments.

A gas analyzer model UMS, provided by Prevac, and a thermoconductivity detector (Gow-Mac) have both been used as downstream detectors (DSD). The gas analyzer contains a mass spectrometer (SRS model RGA 200) and a vacuum system combining a diaphragm pump (Vacuumbrand, MD1, 1.5 hPa) and a turbomolecular pump (Leybold Vacuum, Turbovac TW 70 H). The flow rate of gases was controlled by two Bronkhorst mass flow controllers.

Pressure meter was provided by ZACH Metalchem (Poland) and consisted of a piezoresistive transmitter NPX (Peltron, Poland; range $0.4 - 1$ bar) and a control panel enabling the signal to be recorded continuously into a file.

**Procedures.** All measurements were performed in a continuous sorption mode. The temperature was kept at 75 °C to 77 °C for the sorption of deuterium and a higher temperature range, from 105 °C to 107 °C was used for the sorption of protium. The same Pd sample was used for a succession of days, and a daily sequence comprised of up to five sorption-desorption cycles. A sample ranging from 0.2 g to 0.4 g of palladium powder, depending on granularity, was required to fill the calorimetric cell of 0.15 cm$^3$. The cell was then sealed with a top and bottom gas inlet and outlet, and the sample was thermally equilibrated in a flow of either of the following inert carriers: He, Ne, Ar, Kr, or $N_2$. The outlet was connected to a thermoconductivity detector (TCD) and subsequently, via a stainless steel capillary tube (1m long, 0.12 mm in diameter), to a mass spectrometer (MS) equipped with a diaphragm pump and a turbomolecular pump. Occasionally, a piezoresistive transmitter would be located between the TCD and the MS. The supply of gases to the inlet was controlled by two mass flow controllers and a system of valves making it possible to form a reaction mixture of hydrogen ($H_2$ or $D_2$) and inert gas (He, Ne, Ar, Kr, or $N_2$) at various proportions and to set up flow rates of both the mixture and the inert. At the time of inert passing, when no reaction takes place in the cell, this arrangement makes it possible for the system pressure to be effectively controlled by setting up an inert flow rate while the system is being continuously pumped down. During the sorption however, when a substantial fraction of hydrogen is being removed from the gas phase as it passes through the Pd sorbent, the total pressure can no longer be controlled to a set point, and in fact it is more a function of the reaction rate rather than a function of the flow rate. In some of the experiments, the pressure within the system was monitored continuously with piezoresistive transmitter and recorded concurrently with the calorimetric signal.

A usual experimental procedure was as follows. After reaching thermal equilibrium with the inert gas the sample is ready for a sorption measurement. At this point the flow of inert is lowered in order to decrease the system pressure to around 550 hPa which takes around 20 minutes. The inert is then replaced by a flow of a reaction mixture of hydrogen with the same inert, effectively beginning the process of sorption. Following the admission of reaction mixture, an exothermic peak is recorded by the microcalorimeter and the changes in hydrogen concentration in the effluent from the cell are measured by the TCD and MS. The end of the process is signaled by cessation of heat evolution, as well as a plateau being reached in the curve of the TCD downstream detector, indicating no more uptake of hydrogen by the sample from the reaction mixture. At this point, returning to the pure inert flow initiates desorption from the sample during which an endothermic peak is recorded that concludes a sorption-desorption cycle. The end of the desorption is again signaled by a cessation of heat evolution and by readings from the TCD and MS. With the desorption completed the system is ready for the next cycle. It takes about 25 to 40 min for an average sorption to saturate the Pd sample with $H_2$ or $D_2$, and afterwards around 1.5 h to 2 h for desorption to occur in the flow of pure inert.

Using different temperatures for each isotope was dictated by the need of conducting the sorption experiments within a short range reversibility regime, which means that the duration of both processes, namely sorption and the subsequent desorption, should be comparable, so that their respective thermal effects can be measured with equal precision. For both isotopes this requires certain temperatures higher than the room temperature (RT) at which the desorption is a prolonged process that takes more than 20 h to complete, whereas a typical sorption needs ca. 0.5 h at ambient pressure and temperature. However, it was not possible to find a common temperature setting at which both isotopes would be sorbed reversibly in the above sense. At the temperature of 107 °C, at which the sorption of $H_2$ proceeds reversibly[1], deuterium practically fails to be sorbed by palladium powder, yielding only a very slight thermal effect. In turn, at 75 °C, the highest temperature at which $D_2$ sorbs both substantially and reversibly, the sorption of protium is no longer reversible, that is, the prolonged duration of the desorption makes it practically impossible to measure the accompanying heat evolution precisely. Hence the experiments with protium and deuterium were carried out at different temperatures, on recognizing that the criterion of reversibility is more essential for comparison of reaction courses for both isotopes than assuring the same process temperature.

Calorimetric calibrations (pulses of 30 mW power and 300 s duration) have been made *in situ* following each sorption and desorption. For each sorption peak, the height of its immediately following calibration peak was used as CF to calculate the integrated heat. However, using a single CF for desorption peaks yielded integrated heats that are inconsistent with those obtained for the sorption, on average by ca. 10%. As a remedy, a method using what may be a called dynamic calibration factor (DCF) has been developed (cf. Appendix 1 in the Supplementary Material[13]) and consequently applied in this work for calculating the heats of desorption.

## 3. Results and discussion

**An overview**. The nature of the inert gas (IG) admixed with hydrogen prior to its contact with Pd affects profoundly both the amplitude and the frequency in the ensuing oscillatory sorption process. Figures 1 and 2 demonstrate variations of heat evolution accompanying the sorption of respectively $H_2$ and $D_2$ isotope in Pd powder as a function of the kind of inert gas admixtures, He, Ne, Ar, Kr or $N_2$. Other than the inert, all the controlled reaction parameters were set the same for each isotope. It can be seen that the oscillation amplitudes increase conspicuously on going from He to Kr. In the same order, the maximum rate of heat production reached in the course of sorption also increases moving down the periodic table from He to Kr. Accordingly, in Figures 1a-e for $H_2$ the maxima are: 26.6, 40.7, 48.8, 60.2 and 54.5 mW, respectively for He, Ne, Ar, Kr and $N_2$, and in Figures 2a-e for $D_2$ they are 39.3, 51.7, 53.8, 60.3 and 41.8 mW. Thus for both $H_2$ and $D_2$ the oscillations induced by heavier inert gases are more pronounced. This "inert effect" was found to be rather typical, if not always that conspicuous, and was confirmed in a number of similar groups of experiments.

**Effect on frequency.** Variations in oscillation frequencies can be established by analyzing the Fourier transform of the calorimetric time series. Figure 2f compares the power spectra of the time series for $D_2$ shown in Figures 2a-e. It can be seen from the positions of the peaks that the frequencies decrease in passing from He to Kr. This trend can be further verified by comparing a total of 113 experiments involving both noble gases and nitrogen as admixtures to either isotope. The frequencies obtained in these experiments are listed in Table 1. The data are arranged in such a way that each row represents a group of up to five frequency values, respectively obtained for He, Ne, Ar, Kr and $N_2$, under a specific set of reaction conditions, the latter are listed at the right hand side of the table. Apart from indicating which isotope has been used, $H_2$ or $D_2$, the reaction conditions also include parameters such as the flow rate, the partial pressure of hydrogen or of deuterium, the size of the Pd powder grains, and the details of experimental setup, in this case whether the pressure transducer was installed or not. It can be seen, that within each row in Table 1 the frequencies of the oscillations decrease in the order He > Ne ≥ Ar > Kr. Rows no 16 and 1 correspond to Figures 1 and 2 respectively.

The values given in brackets in Table 1 are the results of repeated measurements. In order to check reproducibility, these pairs of data (14 pairs in total) were used to calculate their normalized absolute differences, yielding on average of 1.35 % which seems to indicate a level of reproducibility sufficient to merit a reliable mathematical analysis of the data. The reproducibility of measurements is also illustrated in Figure 1f by showing an example of two successive experiments with $H_2/N_2$

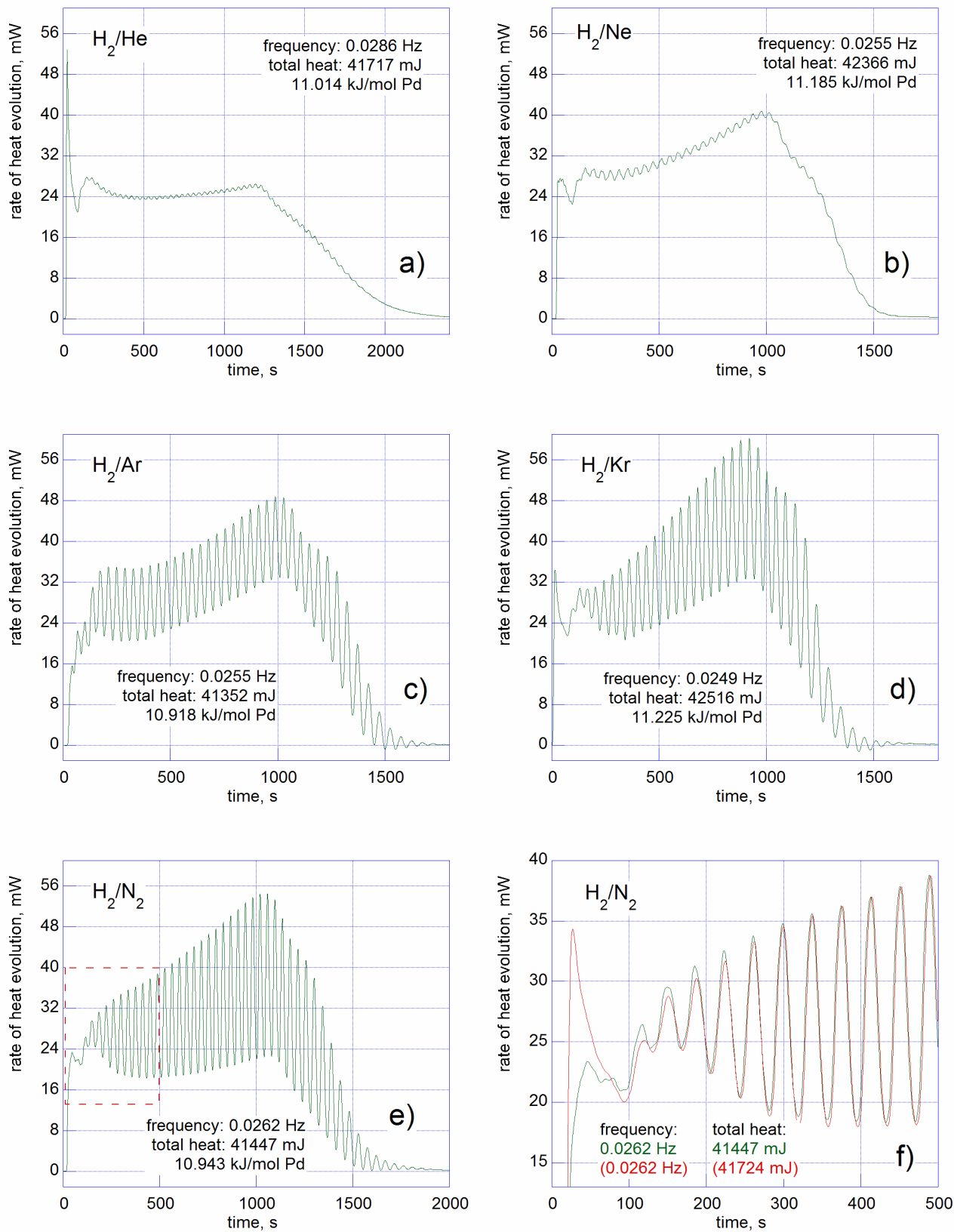

**Figure 1.** The calorimetric time series representing oscillations in sorption of $H_2$ in Pd induced by different inert gases admixed to the flow of $H_2$ prior to its contact with Pd: He (**a**), Ne (**b**), Ar (**c**), Kr (**d**) and $N_2$ (**e**). Panel (**f**) shows repetition of the experiment shown in (**e**). For reaction conditions see group no 16 in Table 1.

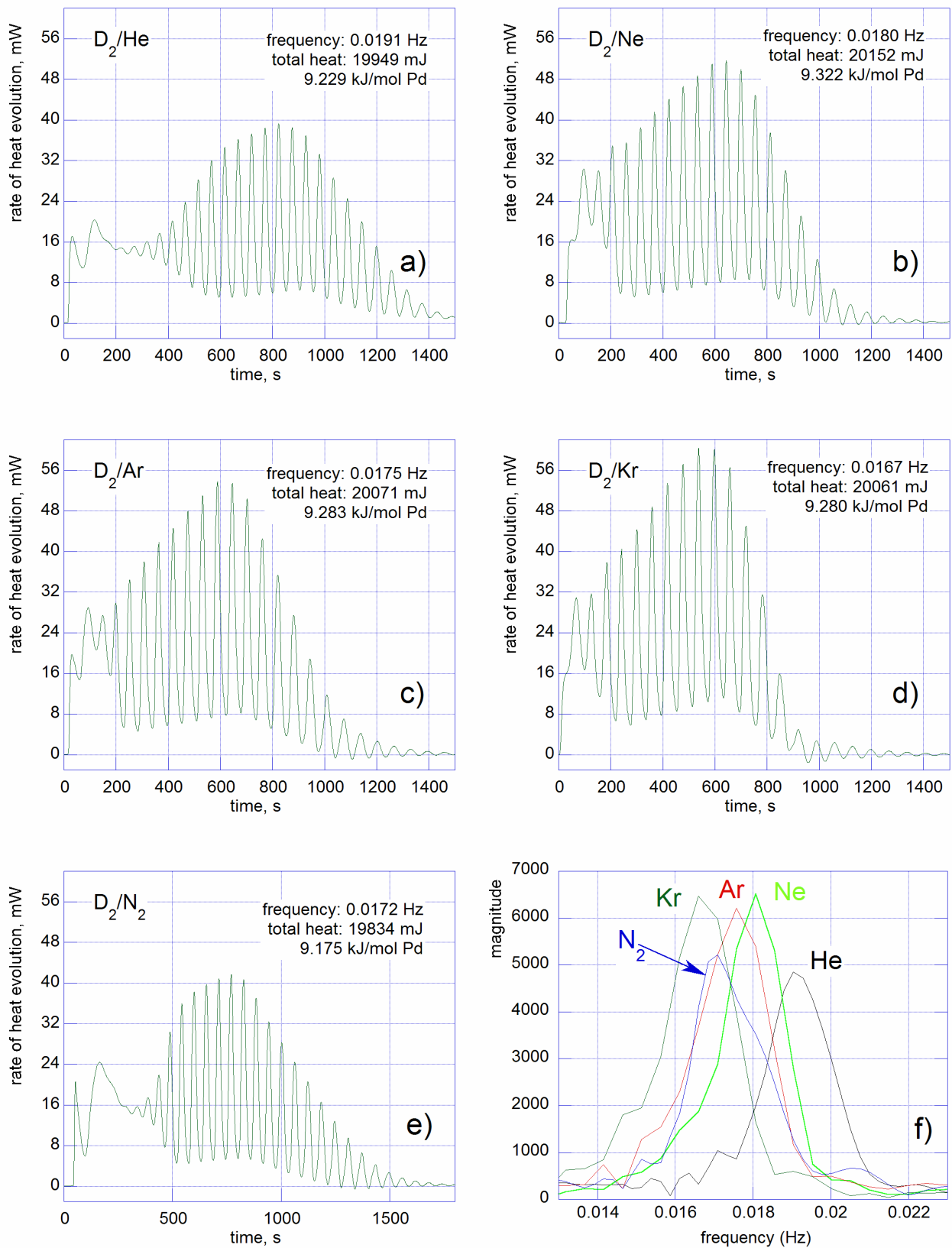

**Figure 2**. The calorimetric time series representing oscillations in sorption of $D_2$ in Pd induced by different inert gas admixed to the flow of $D_2$ prior to its contact with Pd: He (**a**), Ne (**b**), Ar (**c**), Kr (**d**) and $N_2$ (**e**). The corresponding power spectra are presented in panel (**f**). For reaction conditions see group no 1 in Table 1.

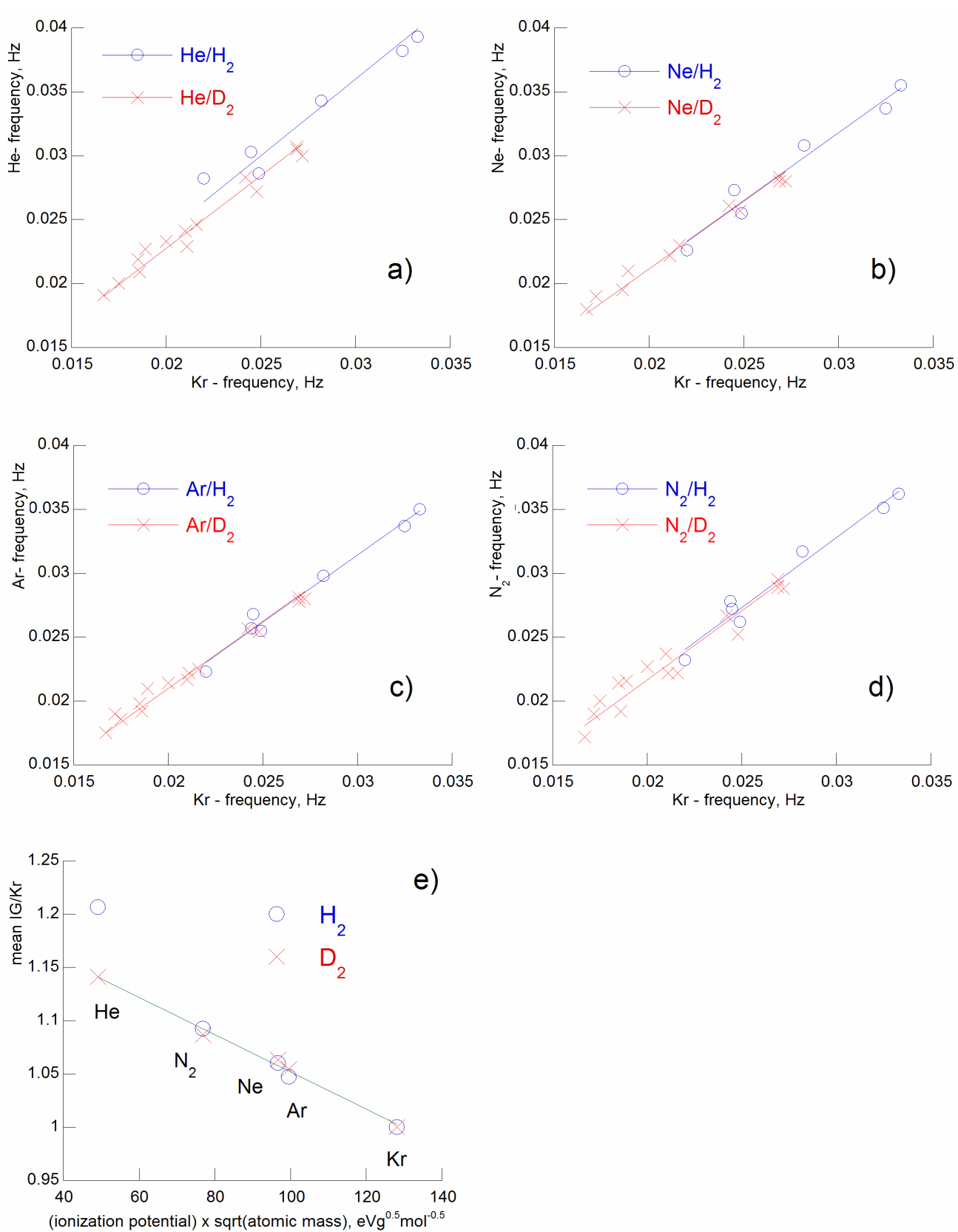

**Figure 3.** (**a**) - (**d**) The frequencies of inert induced oscillations form a straight line when plotted as a function of one another (data listed in Table 1). (**e**) The correlation of the mean normalized frequencies (see text) with the product of the first ionization potential times the square root of atomic mass of inert gases.

**Table 1**. Oscillation frequencies (Hz) in sorption of $H_2$ and $D_2$ in Pd as a function of the kind of inert gas admixture, respectively: He, Ne, Ar, Kr and $N_2$, added to $H_2$ or $D_2$ prior to reaction with Pd. The values in brackets indicate repetitions. The groups 16 and 1 correspond to Figures 1 and 2. The values of flow rate are for standard conditions.

| group No | Oscillation frequencies, Hz | | | | | Isotope of hydrogen | Reaction conditions | | | | |
| --- | --- | --- | --- | --- | --- | --- | --- | --- | --- | --- | --- |
| | He | Ne | Ar | Kr | $N_2$ | | $H_2$ or $D_2$ molar fraction | flow rate, cm³/min | sample mass, g | Pd sample[1] | Press. meter[2] |
| 1 | 0.0191 | 0.0180 | 0.0175 | 0.0167 | 0.0172 (0.0174) | $D_2$ | 0.92 | 2.2 | 0.2300 | powder | 0 |
| 2 | - | 0.0190 | 0.0190 | 0.0172 | 0.0190 | $D_2$ | 0.91 | 1.6 | 0.3032 | sponge | 0 |
| 3 | 0.0200 | - | 0.0186 | 0.0175 | 0.0200 | $D_2$ | 0.91 | 1.6 | 0.2515 | powder | 0 |
| 4 | 0.0219 | - | 0.0198 | 0.0185 | 0.0214 | $D_2$ | 0.90 | 2.0 | 0.2515 | powder | 0 |
| 5 | 0.0209 | 0.0195 | 0.0192 | 0.0186 | 0.0192 | $D_2$ | 0.92 | 2.2 | 0.2300 | powder | 1 |
| 6 | 0.0227 | 0.0210 | 0.0210 | 0.0189 | 0.0216 | $D_2$ | 0.92 | 2.2 | 0.3032 | sponge | 0 |
| 7 | 0.0233 | - | 0.0214 | 0.0200 | 0.0227 | $D_2$ | 0.92 | 2.2 | 0.2515 | powder | 0 |
| 8 | 0.0241 | - | 0.0217 | 0.0210 | 0.0237 | $D_2$ | 0.92 | 2.4 | 0.2515 | powder | 0 |
| 9 | 0.0229 | 0.0222 | 0.0222 | 0.0211 | 0.0222 | $D_2$ | 0.92 | 2.2 | 0.3809 | f. powder | 1 |
| 10 | 0.0246 | 0.0230 | 0.0225 | 0.0216 | 0.0222 | $D_2$ | 0.94 | 3.2 | 0.2300 | powder | 0 |
| 11 | 0.0282 | 0.0226 | 0.0223 | 0.0220 (0.0213) | 0.0232 | $H_2$ | 0.92 | 2.2 | 0.4030 | f. powder | 0 |
| 12 | 0.0283 | 0.0261 | 0.0257 | 0.0242 | 0.0267 | $D_2$ | 0.94 | 3.2 | 0.3032 | sponge | 0 |
| 13 | - | - | 0.0257 | 0.0244 | 0.0278 | $H_2$ | 0.91 | 1.6 | 0.2515 | powder | 0 |
| 14 | 0.0303 (0.0303) | 0.0273 | 0.0268 | 0.0245 (0.0260) | 0.0272 | $H_2$ | 0.92 | 2.2 | 0.3809 | f. powder | 1 |
| 15 | 0.0272 | 0.0257 | 0.0254 (0.0253) | 0.0248 | 0.0252 | $D_2$ | 0.94 | 3.2 | 0.2300 | powder | 1 |
| 16 | 0.0286 | 0.0255 | 0.0255 | 0.0249 | 0.0262 (0.0262) | $H_2$ | 0.92 | 2.2 | 0.4030 | f. powder | 1 |
| 17 | 0.0305 (0.0307) | 0.0280 (0.0283) | 0.0278 (0.0281) | 0.0269 (0.0267) (0.0269) | 0.0295 (0.0289) | $D_2$ | 0.94 | 3.2 | 0.3032 | sponge | 1 |
| 18 | 0.0300 | 0.0280 | 0.0280 | 0.0272 | 0.0288 | $D_2$ | 0.94 | 3.2 | 0.3809 | f. powder | 1 |
| 19 | 0.0343 | 0.0308 | 0.0298 | 0.0282 | 0.0317 | $H_2$ | 0.94 | 3.2 | 0.4030 | f. powder | 0 |
| 20 | 0.0382 (0.0382) | 0.0337 | 0.0337 | 0.0325 | 0.0351 | $H_2$ | 0.94 | 3.2 | 0.4030 | f. powder | 1 |
| 21 | 0.0393 | 0.0355 | 0.0350 | 0.0333 (0.0337) | 0.0362 (0.0367) | $H_2$ | 0.94 | 3.2 | 0.3809 | f. powder | 1 |

1) powder: particle size 0.25 – 2.36 mm; f. powder (fine powder): particle size less than 75 μm; sponge: p article size 44 μm – 149 μm (The same volume of the Pd sample was always used to fill the calorimetric cell; the different sample masses result from various bulk densities of the Pd powders.)
2) pressure transducer installed: (1); without pressure metering: (0)

under the same reaction conditions. One of these time series has been already shown in the previous panel (cf. Figure 1e), with the dashed rectangle marking the enlarged area, which is represented in Figure 1f. It can be seen that nearly exact oscillations have developed in both experiments, in spite of the initially different incubation stages apparent during the first 150 s.

**Analysis of frequency variations.** Each of the 21 rows in Table 1 differs by at least one aspect of the reaction conditions. Consequently, the columns headed respectively as He, Ne, Ar, Kr and $N_2$ each contain a dataset of up to 21 values of frequencies $f_{IG}$ (with IG = He, Ne, Ar, Kr or $N_2$) for the same inert gas at various conditions. Comparing these datasets (columns) is not straightforward, since the frequencies obtained at different conditions cannot be simply averaged within an individual dataset (column). However, it has been found that they can be normalized, and the Kr dataset has been selected as the basis for normalization. In this procedure, the frequencies for He, Ne, Ar and $N_2$ are respectively divided by the corresponding values for Kr (row by row), yielding the ratios of $f_{IG}/f_{Kr}$, that is, normalized frequencies, of which the arithmetic mean values can then be calculated separately for $H_2$ and $D_2$. To confirm the validity of this procedure, Figure 3a-d shows that it is possible to represent the frequency values for He, Ne, Ar and $N_2$ each as a linear function of the Kr frequency. All these straight lines turned out to pass through, or very close to, zero and so they can be represented as a function of the type $f_{IG} = a_{IG}/f_{Kr}$. Therefore, mathematically, the arithmetic means of $f_{IG}/f_{Kr}$ ratios correspond to the $a_{IG}$ parameters, that is, to the slopes of the straight lines represented in Figures 3a-d. Table 2 shows the values so determined: the mean normalized frequencies for $H_2$ and $D_2$. Clearly, the data in Table 2 confirm the trend of oscillation frequency to decrease in the order He, Ne, Ar, Kr. It can also be concluded that, except for helium, there are no isotope effects; that is, the values of $f_{IG}/f_{Kr}$ are practically the same for $H_2$ and $D_2$.

**Table 2.** The mean normalized frequencies for $H_2$ and $D_2$.

|       | He/Kr  | Ne/Kr  | Ar/Kr  | $N_2$/Kr | Kr/Kr  |
|-------|--------|--------|--------|----------|--------|
| $H_2$ | 1.2065 | 1.0601 | 1.0471 | 1.0925   | 1.0000 |
| $D_2$ | 1.1412 | 1.0633 | 1.0547 | 1.0868   | 1.0000 |

**Empirical correlation of frequencies.** It would be of interest to find a functional relation between the mean normalized frequencies and the physicochemical properties representing the nature of the inert gases inducing the oscillatory behavior. However, a surprising feature of the data in Table 2 is the closeness of the values for Ne and Ar. This seems to be unusual when compared to various physical properties of noble gases that rather tend to change smoothly; that is, with relatively similar increase (or decrease) from one gas to the next, moving down the periodic table. In fact, an attempt to find a correlation of the mean normalized frequencies with any single set of the inert gas physicochemical characteristics has failed. Instead, it has been found empirically, that so determined means correlate with a product combining two distinct atomic properties: the first ionization potential (FIP) times the square root of the atomic mass (molecular mass for $N_2$), sqrt($M$). The values of the FIP for the inert gases (IG) He, Ne, Ar, Kr and $N_2$ are respectively: 24.587, 21.564, 15.759, 13.999, 15.580 eV.[14] Correspondingly, the values of the product, $FIP_{IG} * sqrt(M_{IG})$, are respectively: 49.174, 96.678, 99.544, 128.15, 82.420 $eVg^{0.5}mol^{-0.5}$. The linear correlation is represented in Figure 3e. It can be seen that the mean values of $f_{NG}/f_{Kr}$ decrease proportionally to $FIP_{IG} * sqrt(M_{IG})$, except for the mean $f_{He}/f_{Kr}$ for $H_2$, an apparent outlier.

Effectively, the oscillations appear to be enhanced jointly by heavier inert gas atoms (molecules) and by how easily the inert gas can be ionized. It is well established that noble gases can be weakly adsorbed on the Pd surface,[10,15] owing to the dispersion forces of which the magnitude is a function of the first ionization potential of the adsorbate atom.[16] The existence of an empirical relation between the oscillation frequencies and the FIP (cf. Figure 3e) may therefore point to the dispersion forces and, by the same token, to a weak adsorption of the inert gases as being involved in a mechanism responsible for triggering oscillations. The adsorbed atoms of noble gases are polarized and according to DFT calculations the strength of their induced dipole moment as well as the adsorption energy both increase in passing from He to Kr.[10] Hence the heavier species are adsorbed more strongly, and again, this seems to be in agreement with the empirical relation in Figure 3e, which shows that the oscillations are enhanced by heavier inert gases.

The departure of the He/$H_2$ system from the correlation shown in Figure 3e may possibly be related to a peculiar mechanism of the H-He collisions, experimentally found to be unique in the sense that the resultant so-called collision-induced dipole moment in the H atom has a negative value, in contrast to the H-Ne, H-Ar and H-N$_2$ collisions.[17-20] The latter result in either a close to zero (for $N_2$)[17] or positive (for Ne, Ar)[20] direction of electric dipole in hydrogen for the entire energy range from 1 eV to 25 eV. The negative value of the dipole moment means that after a collision the electron is riding in front of the proton rather than lagging behind it. Possibly, the direction of the dipole moment may affect the rate of recombination of atomic hydrogen. The recombination is strongly exothermic and so it may have an impact on the overall rate of heat evolution during the sorption process. It is well established, that hydrogen desorbing from palladium hydride contains a fraction of atomic hydrogen, apart from the $H_2$ molecules.[21] Thus shortly after desorption from Pd these hydrogen atoms may undergo collisions with the inert gas species which thus may alter the recombination rate. Arguably, it would be of interest to consider also the data for collisions of deuterium with noble gases, in particular a value of dipole moment induced by the He-D collisions in D atoms would be revealing. In fact, it might be a crucial test; since the He/$D_2$ system does not depart from the correlation in Figure 3e, so a positive value of dipole moment in D atoms would be in agreement with the proposed explanation. A negative value, on the other hand, would amount to its rejection. To date, however, our literature search for data on dipole moments in the He-D collisions has been unsuccessful.

**Invariance of the thermal effect with respect to inert gas.** Integration of the calorimetric time series yields the total amount of heat that is evolved in a sorption experiment. The integrated heats for the oscillatory sorption of $H_2$ and $D_2$ with various admixtures are respectively given in the appropriate panels in Figures 1 and 2. Comparing the heats of sorption in Figure 1, it is remarkable that in spite of the conspicuous variations between the calorimetric time series, the integrated areas under these curves are the same with a scatter of 2.8 % (between 41352 mJ and 42516 mJ). Table 3 lists the thermal effects for a total of 50 sorption-desorption cycles, using either $H_2$ (24 cycles) or $D_2$ (26 cycles), with various inert admixtures, for reaction with Pd. Columns 3 and 4 of Table 3 present the total heats of sorption and desorption, while columns 6 and 7 show the same data represented as heats per mol of Pd (taking the atomic mass for Pd to be 106.4 g/mol). The average molar heat of $H_2$ sorption for the experiments represented in Table 3 is 11.387 kJ/mol Pd. Approximating the H/Pd ratio to be 0.65 at 106 °C at normal pressure[22,23] the heat of sorption per mol of hydrogen is 35.0 kJ/mol $H_2$, which is similar to the value of 35.4 kJ/mol $H_2$ at 80 °C reported previously.[24,25] It also indicates that the sample has been saturated with hydrogen, or very close to saturation, in each experiment represented in Table 3. Likewise for deuterium, Figure 2 shows that the integrated areas of the calorimetric time series are the same for all inerts, ranging from 19834 mJ to 20152 mJ (scatter of 1.0 %). The average molar heat of $D_2$ sorption for the experiments represented in Table 3 is 9.4355 kJ/mol Pd. Taking a D/Pd ratio of 0.62 at 75 °C at 1000 mbar[23] we obtain 30.4 kJ/mol $D_2$. The difference between the $H_2$ sorption (35.0 kJ/mol $H_2$; cf. previous paragraph) is likely to be entirely due to the isotope effect between $H_2$ and $D_2$. Such an isotope difference (4.6 kJ/mol) seems in good agreement with the calorimetric data reported for the sorption of both isotopes in palladium at RT by Sacamoto at al.[26] that are 37.2 kJ/mol for $H_2$ and 34.2 kJ/mol for $D_2$ respectively, and by Flanagan at al.[27], 38.2 kJ/mol for $H_2$ and 34.6 kJ/mol for $D_2$ respectively.

**Equivalence of heats of sorption and desorption.** Great care has been taken to ensure that within each sorption-desorption cycle the thermal effects measured match each other; that is, the respective absolute values of the heats evolved in sorption and desorption are the same within experimental error, since this equivalence attests to the reversibility of sorption. However, evaluation of the heat of desorption turned out to require a novel calibration procedure, since recognizably the calibration factor was changing with the progress of desorption, due to the decreasing content of hydrogen in the effluent gas. As a result, using the standard calibration method, that is, relying on a single calibration factor, turned out to lead to a 10 % discrepancy between the sorption and desorption thermal effects. A new method employing the so-called dynamic calibration factor (DCF) has been devised and the rationale behind this improved calibration procedure has been given in Appendix 1 in the Supplementary Material[13].

Consequently, Table 3 has the molar heats of desorption calculated using the DCF listed in column 7, next to the molar heats of sorption listed in column 6. The difference between the two was found to be statistically insignificant for the case of $H_2$. Nevertheless, a small bias towards the heats of desorption has been revealed for $D_2$. We argue, however, that this is rather a result of a certain small, systematic error in our calibration procedure (cf.

Appendix 1 in the Supplementary Material[13]), not an indication of irreversibility. The size of this bias can be appraised from Figure 4 which shows the molar heats of sorption and desorption (columns 6 and 7 in Table 3) plotted as a function of the molar fraction of $H_2$ or $D_2$ in the reaction mixture. There is a slight dependence of the thermal effects on the molar fraction of hydrogen in a reaction mixture, similar for both protium and deuterium, no doubt a result of the sorption capacity of palladium being a function of the hydrogen partial pressure in the gas phase. The line representing the molar heats of desorption for $D_2$ lies consistently above the one corresponding to the sorption, but the error is no more than 2 %. Generally, therefore, it is possible to conclude that the thermal effects of sorption and desorption alike are both independent of the kind of inert gas used as an admixture to induce the oscillatory sorption with either isotope. However we never observed any oscillations during desorption.

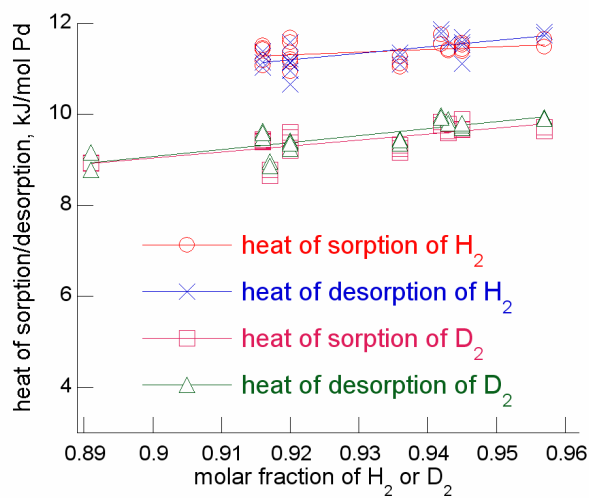

**Figure 4.** Dependence of the heats of sorption and desorption on the molar fraction of $H_2$ and $D_2$ (data listed in Table 3).

**Error estimation.** The invariance of thermal effect of sorption and desorption makes it possible to use the data collected in Table 3 to estimate experimental error in determination of the heat evolution. Since the range of the hydrogen molar fractions used in the experiments is very narrow, from 0.89 to 0.95, so the dependence of the thermal effects on the $H_2(D_2)$ molar fraction can be approximated linearly within this interval. Therefore, the linear fits represented in Figure 4 makes it possible to estimate the expected values of the thermal effects as a function of any value of the $H_2(D_2)$ molar fraction within the same range. The difference of any experimental value from such estimate represents the distance from the data point to the straight line best fitting the data. Arguably, such distance reflects an individual experimental error. Consequently, the average relative error in the heat evolution measurement can be expressed as arithmetic mean of relative absolute differences (RAD) from estimated values of the data in Table 3.

**Table 3.** The thermal effects of sorption and desorption of protium or deuterium admixed with ca. 10 % vol. of either of the following inert gases: He, Ne, Ar, Kr, and $N_2$, prior to contact with palladium.

| No. | composition of reaction mixture | sorption thermal effect, mJ | desorption thermal effect, mJ | sample mass, g | heat of sorption, kJ/mol Pd | heat of desorption, kJ/mol Pd | flow rate, $cm^3$/min | $H_2$ or $D_2$ molar fraction |
|---|---|---|---|---|---|---|---|---|
| 1 | $H_2$ + He | 43042 | -42217 | 0.4030 | 11.364 | -11.146 | 2.2 | 0.920 |
| 2 | $H_2$ + He | 42269 | -42156 | 0.4030 | 11.160 | -11.130 | 2.2 | 0.920 |
| 3 | $H_2$ + He | 43866 | -43890 | 0.4030 | 11.581 | -11.588 | 3.2 | 0.920 |
| 4 | $H_2$ + He | 43153 | -43875 | 0.4030 | 11.393 | -11.584 | 3.2 | 0.945 |
| 5 | $H_2$ + He | 43686 | -43694 | 0.4030 | 11.534 | -11.536 | 3.2 | 0.945 |
| 6 | $H_2$ + Kr | 43844 | -42126 | 0.4030 | 11.576 | -11.122 | 3.2 | 0.945 |
| 7 | $H_2$ + Kr | 43322 | -44304 | 0.4030 | 11.438 | -11.697 | 3.2 | 0.945 |
| 8 | $H_2$ + Kr | 42516 | -42514 | 0.4030 | 11.225 | -11.225 | 2.2 | 0.920 |
| 9 | $H_2$ + Kr | 41472 | -40411 | 0.4030 | 10.949 | -10.669 | 2.2 | 0.920 |
| 10 | $H_2$ + Kr | 44255 | -41499 | 0.4030 | 11.684 | -10.957 | 2.2 | 0.920 |
| 11 | $H_2$ + Ne | 43310 | -43149 | 0.4030 | 11.435 | -11.392 | 2.2 | 0.916 |
| 12 | $H_2$ + Ne | 44527 | -44955 | 0.4030 | 11.756 | -11.869 | 3.2 | 0.942 |
| 13 | $H_2$ + Ne | 43707 | -44630 | 0.4030 | 11.539 | -11.783 | 3.2 | 0.942 |
| 14 | $H_2$ + Ne | 42366 | -43151 | 0.4030 | 11.185 | -11.393 | 2.2 | 0.916 |
| 15 | $H_2$ + Ar | 43425 | -42737 | 0.4030 | 11.465 | -11.283 | 2.2 | 0.916 |
| 16 | $H_2$ + Ar | 43209 | -43768 | 0.4030 | 11.408 | -11.556 | 3.2 | 0.943 |
| 17 | $H_2$ + Ar | 41951 | -42228 | 0.4030 | 11.076 | -11.149 | 2.2 | 0.916 |
| 18 | $H_2$ + $N_2$ | 42686 | -43006 | 0.4030 | 11.270 | -11.355 | 2.1 | 0.936 |
| 19 | $H_2$ + $N_2$ | 42093 | -42732 | 0.4030 | 11.113 | -11.282 | 2.1 | 0.936 |
| 20 | $H_2$ + $N_2$ | 44150 | -44762 | 0.4030 | 11.657 | -11.818 | 3.1 | 0.957 |
| 21 | $H_2$ + $N_2$ | 43537 | -44476 | 0.4030 | 11.495 | -11.742 | 3.1 | 0.957 |
| 22 | $H_2$ + $N_2$ | 41831 | -42041 | 0.4030 | 11.044 | -11.100 | 2.1 | 0.936 |
| 23 | $H_2$ + Ar | 43607 | -41800 | 0.4030 | 11.513 | -11.036 | 2.2 | 0.916 |
| 24 | $H_2$ + Ar | 43274 | -43787 | 0.4030 | 11.425 | -11.561 | 3.2 | 0.943 |
| 25 | $D_2$ + $N_2$ | 19993 | -20467 | 0.2300 | 9.2489 | -9.4683 | 2.1 | 0.936 |
| 26 | $D_2$ + $N_2$ | 21008 | -21486 | 0.2300 | 9.7183 | -9.9396 | 3.1 | 0.957 |
| 27 | $D_2$ + $N_2$ | 19808 | -20257 | 0.2300 | 9.1635 | -9.3709 | 2.1 | 0.936 |
| 28 | $D_2$ + $N_2$ | 20147 | -20403 | 0.2300 | 9.3201 | -9.4386 | 2.1 | 0.936 |
| 29 | $D_2$ + $N_2$ | 20839 | -21428 | 0.2300 | 9.6402 | -9.9126 | 3.1 | 0.957 |
| 30 | $D_2$ + $N_2$ | 18976 | -19373 | 0.2300 | 8.7783 | -8.9622 | 1.7 | 0.917 |
| 31 | $D_2$ + $N_2$ | 18720 | -19177 | 0.2300 | 8.6598 | -8.8713 | 1.7 | 0.917 |
| 32 | $D_2$ + Ar | 19336 | -18994 | 0.2300 | 8.9449 | -8.7868 | 1.8 | 0.891 |
| 33 | $D_2$ + Ar | 20443 | -20523 | 0.2300 | 9.4572 | -9.4941 | 2.2 | 0.916 |
| 34 | $D_2$ + Ar | 21179 | -21398 | 0.2300 | 9.7974 | -9.8987 | 3.2 | 0.943 |
| 35 | $D_2$ + Ar | 20835 | | 0.2300 | 9.6383 | | 3.2 | 0.943 |
| 36 | $D_2$ + Ar | 20352 | -20521 | 0.2300 | 9.4151 | -9.4931 | 2.2 | 0.916 |
| 37 | $D_2$ + Ar | 19288 | -19817 | 0.2300 | 8.9227 | -9.1673 | 1.8 | 0.891 |
| 38 | $D_2$ + Ar | 20727 | -21210 | 0.2300 | 9.5885 | -9.8119 | 3.2 | 0.943 |
| 39 | $D_2$ + Ne | 21266 | -21579 | 0.2300 | 9.8381 | -9.9824 | 3.2 | 0.942 |

| 40 | $D_2$ + Ne | 20316 | -20824 | 0.2300 | 9.3985 | -9.6335 | 2.2 | 0.916 |
| 41 | $D_2$ + Ne | 20382 | -20744 | 0.2300 | 9.4288 | -9.5963 | 2.2 | 0.916 |
| 42 | $D_2$ + Ne | 21114 | -21480 | 0.2300 | 9.7676 | -9.9370 | 3.2 | 0.942 |
| 43 | $D_2$ + Kr | 20755 | -20377 | 0.2300 | 9.6014 | -9.4264 | 2.2 | 0.920 |
| 44 | $D_2$ + Kr | 20931 | -20980 | 0.2300 | 9.6828 | -9.7054 | 3.2 | 0.945 |
| 45 | $D_2$ + Kr | 21372 | -21100 | 0.2300 | 9.8870 | -9.7612 | 3.2 | 0.945 |
| 46 | $D_2$ + Kr | 20514 | -19997 | 0.2300 | 9.4899 | -9.2509 | 2.2 | 0.920 |
| 47 | $D_2$ + He | 20291 | -20326 | 0.2300 | 9.3867 | -9.4029 | 2.2 | 0.920 |
| 48 | $D_2$ + He | 20912 | -21199 | 0.2300 | 9.6742 | -9.8070 | 3.2 | 0.945 |
| 49 | $D_2$ + He | 20885 | -21191 | 0.2300 | 9.6615 | -9.8032 | 3.2 | 0.945 |
| 50 | $D_2$ + He | 19918 | -20240 | 0.2300 | 9.2141 | -9.3633 | 2.2 | 0.920 |

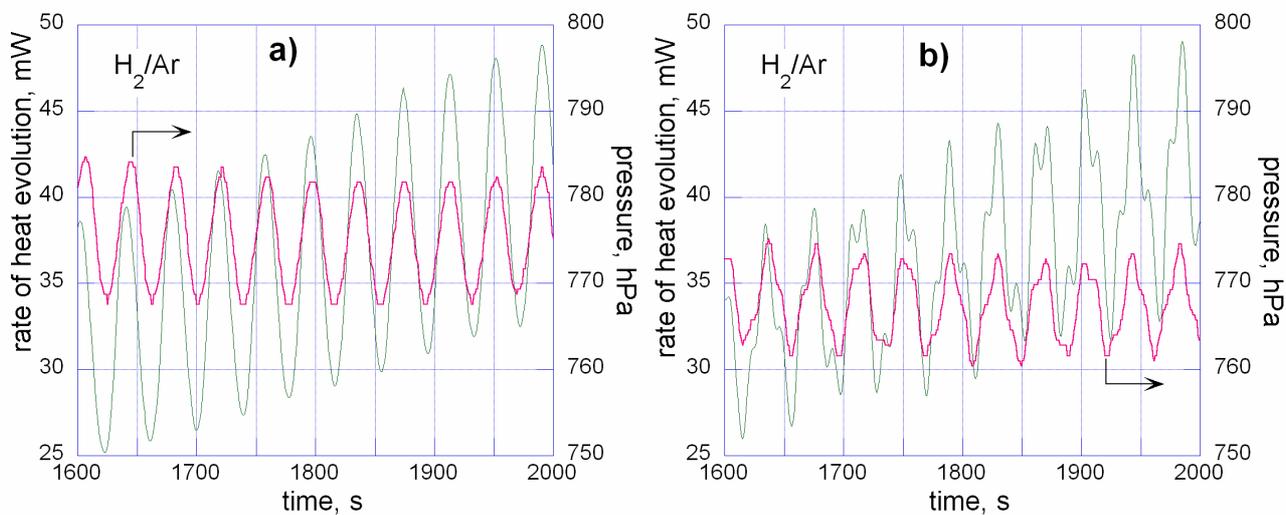

**Figure 5.** The pressure vibrations recorded by piezoelectric transducer downstream the microcalorimetric cell closely correspond to the concurrent thermokinetic oscillations accompanying the reaction of a mixture of $H_2$+Ar with palladium: **a)** periodic oscillations; **b)** quasiperiodic oscillations.

The following formula has been used: RAD = 100 $|x_i - x_{est}|/[0.5(x_i + x_{est})]$, where $x_i$ is an experimental thermal effect, and $x_{est}$ is the estimated value at the same hydrogen molar fraction. Four different $x_{est}$ values have been used, separately for $H_2$ and $D_2$ for sorption and desorption. Consequently, the following mean values for the RADs have been obtained: $H_2$(sorption) 1.42 %; $H_2$(desorption) 1.48 %; $D_2$(sorption) 1.86 %; $D_2$(desorption) 1.64 %. It can be concluded, that for the experiments represented in Table 3, the error in determining the amount of heat evolution is less than 2 %.

**Pressure variations.** In some experiments, the pressure within the system was monitored continuously and found to oscillate in phase with the thermokinetic oscillations. Figures 5 a and b show that the pressure variations exhibits frequencies closely corresponding to those of the thermokinetic counterparts, and a remarkable match between the pressure and the heat evolution time series has been observed even for a case as complex as the quasiperiodic oscillations shown in. Figure 5b. The pressure waves in gaseous hydrogen propagate with a speed roughly four times that of the sound in air,[28] and so the piezoelectric transducer located at a distance of around 1 m downstream the calorimetric cell, can record the pressure changes accompanying the oscillatory sorption with a delay negligible compared to the oscillation period. These are preliminary results and a more full account of the research into pressure oscillations in the sorption of hydrogen in palladium will be published soon. However, the close correspondence of both the pressure and the heat evolution seems to be an important finding, not least because it provides a method other than microcalorimetry for monitoring the oscillations' occurrence during the sorption. It thus brings about the possibility of using various *in situ* techniques to study the oscillating sorption system, while the concurrent monitoring of pressure should be able to provide an accurate account of the oscillations' dynamics.

## 4. Conclusions

Oscillatory heat evolution accompanying the process of sorption of gaseous hydrogen in metallic palladium powder can be induced by an admixture of around 6 % to 10% vol. of an inert gas, such as He, Ne, Ar, Kr and $N_2$ in a reaction mixture with either protium or deuterium. The oscillation frequency always decreases in the order He > Ne > Ar > Kr. The oscillation amplitude increases in the same order. The frequencies for particular inert gases are mutually interdependent, meaning that they form straight lines (passing through zero) when plotted as a function of one another. The mean oscillation frequencies for particular inert gases normalized with respect to Kr have been found to correlate linearly with a product of the inert gas first ionization potential, times the square root of its atomic (molecular) mass. Effectively, the oscillatory rate appears to be enhanced both with the heavier atomic mass of inert gas as well as with its lower ionization energy.

The thermal effect of sorption, on the other hand, is conserved throughout the whole range of the inert gases that have been used as the admixture for either isotope. Moreover, within individual sorption-desorption cycles the heats of the two reactions are, within experimental error, the same, indicative of a total reversibility of the sorption process in Pd. However, no oscillations have been observed during desorption. Thus the use of inert gases as admixtures with either isotope does not alter the thermodynamics of sorption, and hence the oscillations appear to be a kinetic phenomenon.

A remarkable feature of the oscillatory sorption of $H_2$ and $D_2$ in Pd appears to be its very regular and deterministic behavior. By this one means not only the strictly linear dependencies between the oscillation frequencies induced by various inert gas admixtures (cf. Figure 3a-d), but also the very good reproducibility of the results, that can be easily replicated (cf. Table 1; Figure 1f) provided that the reaction conditions are the same. The consistent trend exhibited by oscillation parameters, with their variability following the noble gas order He, Ne, Ar and Kr, as well as the linear dependence of frequencies on the FIP and the sqrt($M_{IG}$) all seem to indicate that the role played by the gas admixtures in the sorption of either isotope in Pd may be far from inert. One may rather suggest that a deterministic interaction involving the inert gas admixture takes place in the Pd/hydrogen system.

**Acknowledgment**. The author is grateful to Microscal Ltd., London, for support and assistance.

## References

[1] E. Lalik, J. Haber, and A. J. Groszek, J. Phys. Chem. C **112**, 18483 (2008).

[2] C. Qi, M. Okumura, T. Akita, and M. Haruta, Appl. Catal. A **263**, 19 (2004).

[3] A. J. Groszek, J. Haber, and E. Lalik, Method for activating a catalyst, British Patent 2430394, (14 May 2008).

[4] Y. Finkelstein, A. Saig, A. Danon, and J. E. Koresh, J. Phys. Chem. B **107**, 9170 (2003).

[5] A. Saig, Y. Finkelstein, A. Danon, and J. E. Koresh, J. Phys. Chem. B **107**, 13414 (2003).

[6] E. Lalik, R. Mirek, J. Rakoczy, and A. Groszek, Catalysis Today **114**, 242 (2006).

[7] M. C. Mittelmeijer-Hazeleger, F. P. Alexandre, A. F. P. Ferreira, and A. Bliek, Langmuir **9**, 3317 (1993).

[8] M. C. Mittelmeijer-Hazeleger, F. P. Alexandre, A. F. P. Ferreira, and A. Bliek, "Adsorption science and technology," in *Proceedings of the Third Pacific Basin Conference*, Kyongju, Korea, 25–29 May 2003, edited by C.-H. Lee (World Scientific Publishing Company, Singapore, 2003).

[9] M. C. Mittelmeijer-Hazeleger, PhD Thesis, University of Amsterdam, 2006.

[10] J. L. F. De Silva and C. Stampfl, Phys. Rev. B **77**, 045041 (2008).

[11] A. J. Groszek, Thermochim. Acta **312**, 133 (1998).

[12] R. Brown and A. J. Groszek, Langmuir **16**, 4207 (2000).

[13] See Supplementary Material below for the rationale of the dynamic calibration factor.

[14] CRC Handbook of Chemistry and Physics, 83rd edition, D. R. Lide Ed., CRC Press 2002, pp. **10**-178 and **10**-185.

[15] A. J. Groszek, E. Lalik, and J. Haber, Appl. Surf. Sci. **252**, 654 (2005).

[16] F. London, Trans. Faraday Soc. **33**, 8 (1937).

[17] R. Krotkov and J. Stone, Phys. Rev. A **22**, 473 (1980).

[18] R. Hippler, O. Plotzke, W. Harbich, H. Madeheim, H. Kleinpoppen and H.O. Lutz, Z. Phys. D **18**, 61 (1991)

[19] R. Hippler, O. Plotzke, W. Harbich, H. Madeheim, H. Kleinpoppen and H.O. Lutz, Phys. Rev. A **43**, 2587 (1991)

[20] B. Siegmann, G.G. Tepehan, R. Hippler, H. Madeheim, H. Kleinpoppen and H.O. Lutz, Z. Phys. D **30**, 223 (1994).

[21] E. Nowicka and R. Duś, Progress in Surface Science **48**, 3 (1995).

[22] W. Betteridge and J. Hope Platinum Metals Rev. **19**, 50 (1975).

[23] R. Lasser and K. –H. Klatt, Phys. Rev. B **28**, 784 (1983).

[24] L. J. Gillespie and F. P. Hall, J. Am. Chem. Soc. **48**, 1207 (1926).

[25] L. L. Jewell and B. H. Davis, Appl. Catal. A **310**, 1 (2006).

[26] Y. Sakamoto, M. Imoto, K. Takai, T. Yanaru, and K. Ohshima, J. Phys.: Condens. Matter **8**, 3229 (1996).

[27] T. B. Flanagan, W. Luo, and J. D. Clewley, Journal of The Less-Common Metals **172-174**, 42 (1991).

[28] CRC Handbook of Chemistry and Physics, 83rd edition, D. R. Lide Ed., CRC Press 2002, p. **14**-42.

**Supplementary Material**

**Appendix 1**

    **Calibration procedure**. A calibration pulse of a controlled power and duration made *in situ* before and/or after sorption provides a reference, that is a calibration peak, of which either height or area can be used as a calibration factor (CF) for calculating the rate of heat evolution for each point of the calorimetric curve. This is a relatively simple procedure provided it can be assumed that a calibration factor remains constant throughout the sorption experiment. In the case of our experimental sorption-desorption cycles, only the sorption part meets this condition. For metallic samples, such as those of the Pd powder, that are highly thermal conductive, the CF is strongly dependent on the thermal conductivity of the gas flowing through the cell, apart from being a function of the thermal conductivity of palladium. Therefore, the CF obtained when pure hydrogen is flowing through the cell is different than the CF for a pure inert, with the CF falling in between for a mixture. It is impossible to introduce calibration pulses while the reaction is in progress, as this would disturb the measurement, and so one has to rely on the CFs obtained from calibration pulses before and after a heat evolution event.

    The actual value of the CF in a flow of gas mixture is obviously a function of the mixture's thermal conductivity, which in turn depends on its concentration, but the latter is a strongly nonlinear dependence. Fortunately, the signal from the thermal conductivity detector, TCD, is analogously a function of the thermal conductivity of the gaseous mixture flowing through it, and so it should be possible to use the TCD signal to evaluate the CF. Indeed, it has been found that the height of a calibration pulse is (approximately) linearly related to the level of the TCD signal when the same gas mixture flows through both the TCD and the calorimetric cell (cf. Figure A1). Figure A2 shows an example of a TCD profile plotted concurrently with the calorimetric curve. There are three 30 mW calibration pulses visible in the calorimetric curve: The first one, at the beginning, is made when a pure inert is flowing through the cell over the palladium sample. The second, in between the sorption and desorption peaks, is measured while a mixture of ca. 90 % of hydrogen and inert is flowing through the cell. Afterwards, however, the content of hydrogen in the effluent decreases gradually with the progress of desorption (cf. the *x* % hydrogen range in Figure A2), reaching negligible amounts when it nears completion. Hence the third calibration pulse, which is made long after the end of the desorption reaction, is supposed to see only a pure inert flowing through the cell. Clearly the middle peak (under 90 % $H_2$) is higher than the other two (in pure inert), in spite of all three pulses being of the same 30 mW power. Using a single CF for the whole cycle led, as it turned out, to serious discrepancies (more than 10%) between the integrated heats of sorption and desorption. In fact, Figure A2 shows that the TCD profile stays approximately at the same level during the entire sorption process, and therefore the middle pulse (the height of its peak) can be used as the calibration for the sorption. In contrast, however, during the desorption reaction the TCD profile rises steadily, and so using a single CF for the entire desorption step must lead to an error.

    In order to minimize this error, the dynamic calibration factor (DCF) was calculated for desorption process. The idea behind the DCF is that as the composition of the reaction mixture flowing through the cell changes with reaction progress, so does the height of the calibration peak. But these are the same changes in composition that actually affect the TCD response. Therefore, in order to calculate the DCF we note, that the following proportion relationship should hold for the situation represented in Figure A2:

$$\frac{L - L(1)}{L(2) - L(1)} = \frac{h(1) - h}{h(1) - h(2)}.$$
(1)

This means that given the heights of the two calibration peaks for the two extreme cases: $h(1)$ for a 90 % hydrogen mixture, and $h(2)$ for a pure inert flowing through the cell, and having recorded a full TCD profile that changes from the level $L(1)$ for the 90 % hydrogen mixture to $L(2)$ for the pure inert (cf. Figures A2 and A3), it is possible to determine a height $h$ of, say, an imaginary calibration pulse coinciding with any level $L$ in the TCD response. This is a consequence of the linear relation between changes in (the level of) the TCD response and in the heights of the calibration peaks. Using the peak height as a calibration factor, so that effectively $h$ = DCF, the profile of DCF can be calculated as follows:

$$\text{DCF} = h(1) - \frac{L[h(1) - h(2)] - L(1)[h(1) - h(2)]}{L(2) - L(1)}.$$
(2)

The resultant DCF time profile is illustrated in Figure A4 together with the corresponding desorption peak.

    Using the formula (2), and the resulting DCF profiles for the desorption peaks, it was possible to obtain approximately the same heats of sorption and desorption for the individual cycles (cf. Table 3 and Figure 4). Indeed, the average relative difference between the heats of sorption and desorption calculated for the 50 cycles represented in Table 3 is only 1.6%, a clear improvement considering the more than 10% discrepancy obtained with a single CF. However, it seems the method of DCF has been somehow more successful when applied for the experiments with protium than it has been with deuterium. The Wilcoxon Rank Sign Test for paired data has been applied separately for the protium and deuterium sorption-desorption cycles shown in Table 3. For the protium-involving cycles, the test yielded $P$ = 0.6033

indicating that the differences between the heats of sorptions and desorptions paired within individual cycles are statistically insignificant. For the deuterium-involving cycles, however, the same test yielded $P = 0.0034$, indicative of a statistically significant bias. This can also be seen in Figure 4 showing that statistically the heats of desorption of deuterium are slightly higher than the heats of its sorption. The difference is on average less than 2%, but it seems to indicate that there is still a systematic error involved in calculating the integrated thermal effects for this group of experiments.

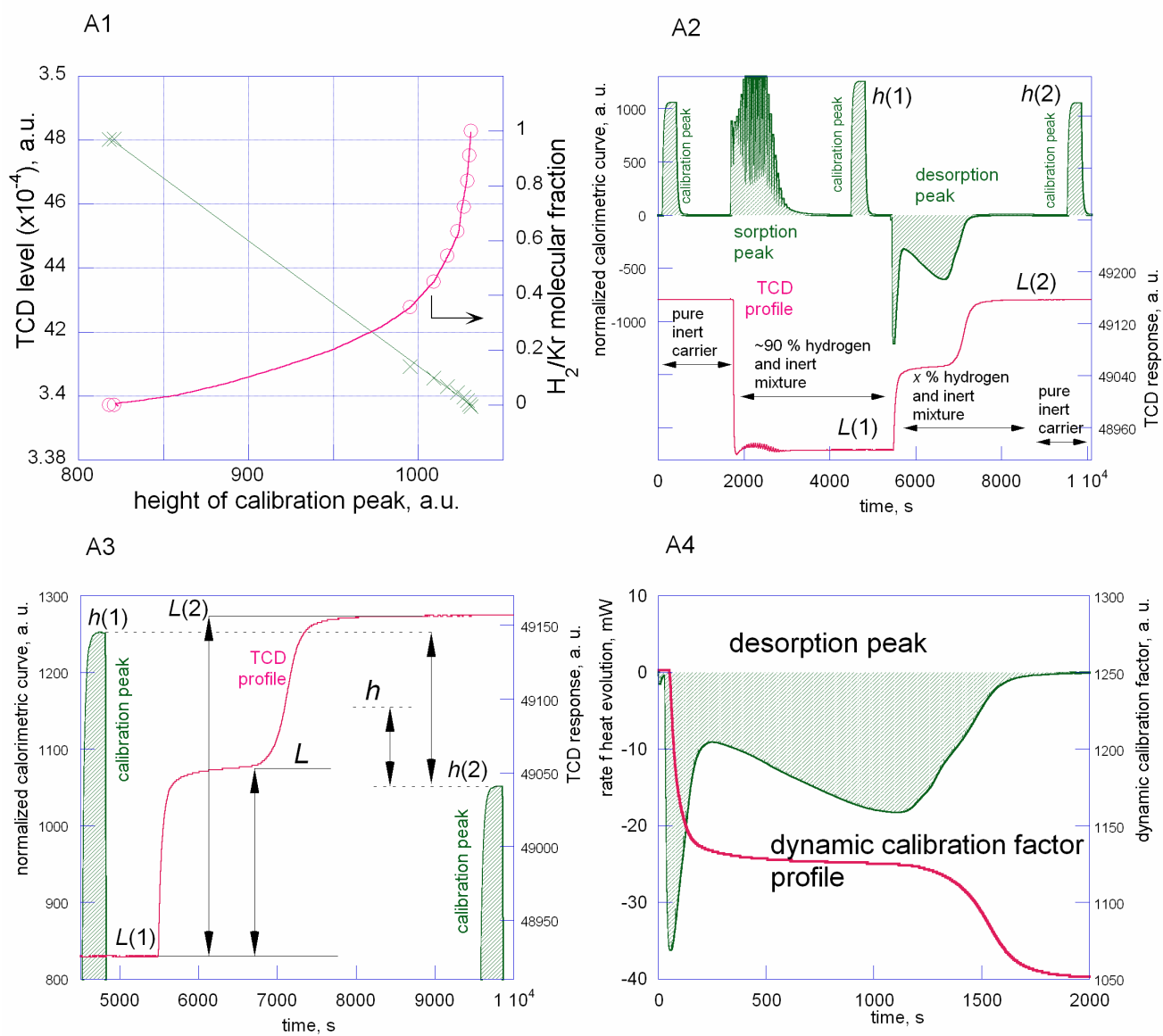

**Figure A1-4.** **(1)** The height of calibration pulse is approximately linearly related to the level of the TCD signal, compared to its dependence on the $H_2$ molar fraction. **(2)** Sorption-desorption cycle shown with the concurrent TCD profile as a function of time. In spite of their different heights, all the three calibration peaks represent identical calibration pulses of 30 mW power and 300 s duration. **(3)** A section of Figure A1 showing the heights $h(1)$ and $h(2)$ of calibration peaks against the TCD profile ranging from $L(1)$ to $L(2)$. The variable $h$ represents a height of a would-be calibration peak, had it been possible to make a calibration pulse at the moment of the TCD profile reaching level $L$. **(4)** A profile of the dynamic calibration factor (DCF) ready to be used for calculating the thermal effects represented by the corresponding desorption peak.